\documentclass[twocolumn,showpacs,floatfix,superscriptaddress,amsmath,amssymb,pre]{revtex4}
\usepackage{mathrsfs}
\usepackage{amssymb}
\usepackage{graphicx}
\usepackage{hyperref}
\usepackage{ulem}
 \usepackage{overpic}
\usepackage{color}
 \usepackage{psfrag}
\usepackage{tabularx}
\usepackage{array}
\usepackage{placeins}
\makeatletter

\begin{document}

\title{Critical exponents of block-block mutual information in one-dimensional infinite lattice systems}

\author{Yan-Wei Dai}
\affiliation{Centre for Modern Physics,
Chongqing University, Chongqing 400044,  China}

\author{Xi-Hao Chen}
\affiliation{ Research Institute for New Materials and Technology, Chongqing University of Arts and Sciences, Chongqing 402160, China
}

\author{Sam Young Cho}
\altaffiliation{E-mail: sycho@cqu.edu.cn}
 \affiliation{Centre for Modern
Physics, Chongqing University, Chongqing 400044,  China}
 \affiliation{Department of
Physics, Chongqing University, Chongqing 400044,  China}

 \author{Huan-Qiang Zhou}
 \affiliation{Centre for Modern
 Physics, Chongqing University, Chongqing 400044, China}
 \affiliation{Department of
Physics, Chongqing University, Chongqing 400044,  China}

 \begin{abstract}
  We study the mutual information between two lattice-blocks in terms of von Neumann entropies
   for one-dimensional infinite lattice systems. Quantum $q$-state Potts model and transverse field spin-$1/2$ XY model are considered numerically by using the infinite matrix product state (iMPS) approach. As a system parameter varies, block-block mutual informations exhibit a singular behavior that enables to identify critical points for quantum phase transition.
  As happens
  with the von Neumann entanglement entropy of a single block,
  at the critical points, the block-block mutual information between the two lattice-blocks of
  $\ell$ contiguous sites equally partitioned in a lattice-block of $2\ell$ contiguous sites
  shows a logarithmic leading behavior, which yields the central charge $c$ of the
  underlying conformal field theory.
  As the separation between the two lattice-blocks increases, the mutual information
  reveals a consistent power-law decaying behavior for various truncation dimensions and lattice-block sizes.
  The critical exponent of block-block mutual information in the thermodynamic limit
  is estimated by extrapolating the exponents of power-law decaying regions
  for finite truncation dimensions.
  For a given lattice-block size $\ell$,
  the critical exponents for the same universality classes
  seem to have very close values each other.
  Whereas
  the critical exponents have different values to a degree of distinction
  for different universality classes.
  As the lattice-block size becomes bigger, the critical exponent becomes smaller.

 \end{abstract}

 \maketitle

 \section{Introduction}
\begin{figure}
 \includegraphics[width=0.65\linewidth]{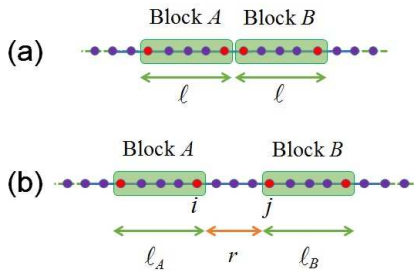}
 \caption{(color online)  Two (a) adjacent
  and (b) disjoint lattice-blocks $A$ and $B$ in one-dimensional infinite lattice systems.
  The block sizes are denoted by $l_A$ and $l_B$, respectively.
  The size of the blocks $l_A$ and $l_B$ are denoted by the number of sites inside the blocks, respectively.
  $r=|i-j|$ is the lattice distance between two blocks.
}
\label{figure1}
 \end{figure}

 As an information of one system about another, correlations quantify a relationship or connection
 between them.
 Correlations have long been a central theme of physical quantities
 in characterizing a unique property of strongly correlated systems in condensed matters.
 Conventional two-point spatial correlation functions have been studied and their scaling behaviors
 have been then used to characterize quantum phases of
 many-body systems \cite{Sachdev,Chaikin}.
 Recently much attentions have been drawn to quantum entanglement that can quantify unique correlations
 present in quantum states.
 Such as entanglement entropy, concurrence, and R\'{e}nyi entropy,
 quantum information theoretical tools have been
 shown to be useful to investigate quantum critical points
 and different phases in strongly correlated systems \cite{Amico}.
 Similar to conventional two-point correlations, in general,
 a correlation between two blocks embedded in a large system may also be considered
 to study a characteristic behavior of the system.
 For a chosen size of blocks,
 block-block correlations
 can be in principle from either classical or quantum origin.
 Not due to entanglement, nontrivial quantum correlations can exist \cite{Dorner}.
 Although correlations can be induced from such different origins
 and specific dominant correlations are not known to characterize the system,
 the mutual information can be used to measure all kinds of correlations of one block
 about the other, i.e.,
 the total amount of classical and quantum correlations between two blocks
 \cite{Adami,Groisman,Schumacher,Eisler}.
 In terms of the von Neumann entropies,
 the mutual information $I(A:B)$
 between two lattice-blocks $A$ and $B$ (see Fig. \ref{figure1})
 can be defined as
\begin{equation}
 I(A:B) = S(A) + S(B) - S (A \cup B),
 \label{Mutual}
\end{equation}
 where $S(\alpha)= - \mathrm{Tr} \rho_\alpha \log_2 \rho_{\alpha}$
 is the von Neumann entropy for the lattice-blocks with $\alpha \in \{A, B, A\cup B\}$.
 To calculate the mutual information between two lattice-blocks in the system,
 the density matrix $\rho_\alpha$
 can be expressed in terms of expectation values of operators in the blocks.
 The elements of the density matrix have the form of generalized correlations functions
 and contain, by definition, all block-site correlations.
 The block-block mutual information in terms of the von Neumann entropy consists
 of a weighted average of generalized correlation functions
 and, in fact, measures the strength of the overall correlation
 between two blocks of sizes $\ell_A$ and $\ell_B$.
 This implies that without knowing a dominant correlation between blocks in the system
 and its corresponding operator,
 the mutual information can capture a characteristic property of the system
 even if hidden or exotic correlations present.
 The two-point pairwise mutual information
 has then been used to study quantum phase transitions \cite{Anfossi,Wolf,Melko,Chen,Singh,Um,Wilms,Huang14,Alcaraz1, Stephan14,Alcaraz2,Alcaraz3,Dai1}.
 A recent study
 shows that similar to two-point spatial correlations,
 two-point pairwise mutual information
 can characterize one-dimensional quantum critical systems by using its critical exponent
 \cite{Dai1}.
 In contrast with two-point pairwise mutual information, criticality of systems has
 also been studied by considering
 the mutual information between the two lattice-blocks
 for bipartite systems ~\cite{Alcaraz1, Stephan14, Alcaraz2,Alcaraz3}.

 In order to more deeply understand mutual information in critical systems,
 it would be interesting to study mutual information between
 lattice-blocks embedded in infinite lattice systems
 in Fig. \ref{figure1}.
 Two consecutive subchains in one-dimensional spin models
 (e.g., see Fig. \ref{figure1} (a))
  exhibit a logarithmical growth of the entanglement with respect to the size of subchains
 at critical points.
 Compared with such two consecutive subchains,
 no correlation between disjoint lattice-blocks (e.g., see Fig. \ref{figure1} (b))
 is expected for a large separation at a transition.
 Then it would be interesting to study (i) how this correlation between disjoint lattice-blocks
 behaves with their
 separation $r$, i.e., how it scales, at a quantum phase transition
 and also (ii) how the size $\ell$ of the lattice-blocks affects the scaling.
 Such blocks in critical systems
 \cite{Caraglio,Furukawa,Calabrese09,Marcovitch09,Wichterich09-1,
 Wichterich09,Casini,Calabrese11,Santos11,Calabrese10,Fagotti10,
 Calabrese12,Fagotti12,Calabrese13,Calabrese13-2,Coser,Nobili,Calabrese18,German20}
 have been considered to investigate
 the entanglement entropy, the entanglement negativity, and the mutual information.
 Especially, the scaling functions of the von Neumann entanglement entropy (mutual information)
 have been studied for the Ising model \cite{Calabrese10,Calabrese11,Calabrese18} and
 for the spin-$1/2$ XXZ model \cite{Furukawa,Calabrese18}.
 Actually, studying their scalings has been found to be a nontrivial task because
 the scaling functions depend not only on the central charge but also on more universal
 information of the conformal field theory \cite{Caraglio,Furukawa,Calabrese09}.

 In this paper we investigate the block-block mutual information
 between two lattice-blocks in infinite-lattice systems by using
 the infinite matrix product state (iMPS) representation  with
 the infinite time-evolving block decimation (iTEBD) method \cite{Vidal03,Vidal07,Su2}.
 To consider various universality classes at critical points,
 we consider quantum $q$-state Potts model and transverse field spin-$1/2$ XY model
 and calculate the von Neuman mutual information for various sizes of lattice-blocks.
 We demonstrate that the block-block mutual information
 can be a useful probe for detecting quantum phase transition.
 The scaling of the block-block mutual information is studied
 at the critical point in the thermodynamic limit.
 The numerical results show that similar to conventional two-point correlations,
 block-block information exhibits power-law decaying behaviors.
 We find that as the size of lattice-block $\ell$ increases,
 the critical exponent decreases.
 The scaling behaviors of groundstate block-block mutual informations
 are discussed in associations with a characterization of critical systems.

  This paper is organized as follows.
  In Sec. II, we briefly introduce one-dimensional $q$-state quantum Potts model and
  numerical iMPS approach. A singular behavior of block-block mutual information
  appears to identify a quantum phase transition.
  As the size of lattice-block increases, block-block mutual information increases.
  In Sec. III, at the critical point,
  such increments of block-block mutual information between two lattice-blocks of $\ell$ contiguous sites equally partitioned in a block of $2\ell$ contiguous sites
  show a logarithmic scaling behavior. Its characteristic numerical scaling coefficient
  relying on $q$
  is discussed in connection with a central charge in terms of block-block mutual information.
  In Sec. IV,
  block-block mutual information is shown to decrease as the separation of blocks increases.
  Its power-law decaying behaviors and the critical exponents are discussed in the thermodynamic limit for transverse field spin-$1/2$
  XY model as well as $q$-state quantum Potts model for various sizes of lattice-blocks.
  A summary and remarks of this work are given in Sec. V.
  In Appendix \ref{appA}, the detailed discussions on central charges from block entanglement entropies are made for one-dimensional quantum $q$-state Potts model.
  Appendix \ref{appB} shows the detailed scaling behaviors of block-block mutual informations
  for the transverse-field spin-$1/2$ XY model.

 \section{one-dimensional $q$-state quantum Potts model and mutual information}

 We consider
  $q$-state quantum Potts model ~\cite{Solyom} with the nearest neighbor interaction
 in a transverse magnetic field $\lambda$.
 The Hamiltonian can be written as
 \begin{equation}
 H = -\sum_{i=1}^{\infty}\sum_{p=1}^{q-1}
   \left[ M_{x,p}^{i}M_{x,q-p}^{i+1}+\lambda M_{z}^{i} \right],
    \label{qHam}
  \end{equation}
 where $M_{x,p}=(M_{x,1})^{p}$
     and the Potts spin matrices $M_{x/z}$ are given as
 \begin{equation}
     M_{x,1}=\left(
     \begin{array}{cc}
     0 & I_{q-1}\\
     1 & 0\\
     \end{array}
     \right)
 ~\mbox{and}~
     M_{z}=\left(
     \begin{array}{cc}
     q-1 & 0\\
     0 & I_{q-1}\\
     \end{array}
     \right).
     \end{equation}
 with the $(q - 1) \times (q - 1)$ identity matrix $I_{q-1}$.
 Due to the spontaneous symmetry breaking with the symmetry group
 $Z_{q}$, $q$-degenerate ground states emerge in the broken symmetry phases.
 It is known that a (dis-)continuous quantum phase transition occurs for ($q > 4$) $q\leq4$
 in the one-dimensional quantum $q$-state Potts model
  \cite{potts1,Baxter,Martin,su}.

\begin{figure}
\includegraphics[width=0.5\textwidth]{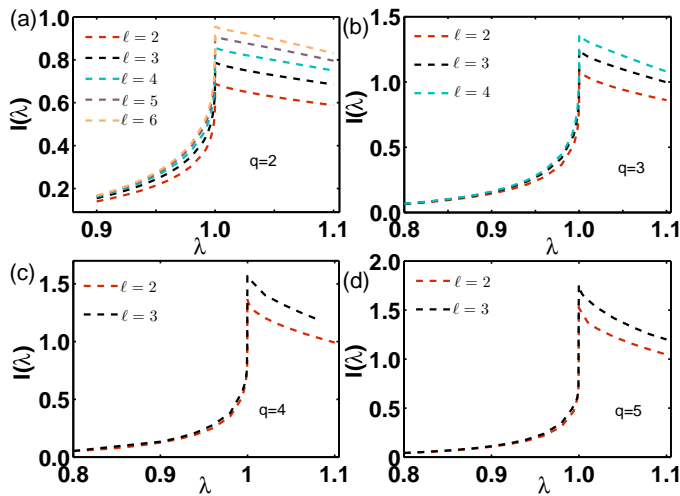}
\caption{(color online) Mutual information $I_q(\lambda)$
 of two adjacent lattice-blocks
 as a function of the transverse field $\lambda$ for
 one-dimensional $q$-state quantum Potts model with various sizes of lattice-blocks $\ell$.
 }
 \label{fig2}
 \end{figure}

 In order to consider an one-dimensional infinite lattice of the system,
 we employ a wave function $\left|\psi\right\rangle$ of Hamiltonian
 in the iMPS representation.
 The iTEBD algorithm
 with the second-order Trotter decomposition leads to
 a numerical ground state $\left|\psi_g\right\rangle$ in the iMPS representation.
 As the initially chosen state approaches to a groundstate, according to a power law,
 the time step is decreased from an initial time step $dt = 0.1$ to $dt = 10^{-6}$.
 Then numerical iMPS wavefunctions for ground states are obtained for the
 truncation dimensions between $\chi = 20$ and $\chi = 150$.
 Actually, in the broken-symmetry phases, randomly chosen several initial states
 can reach different orthogonal groundstates that are degenerate ground states for
 a spontaneous symmetry breaking
 and can be distinguished by using quantum fidelity \cite{su, dai2}.
 Our iMPS approach gives the full description of the groundstate
 in a pure state by the iMPS groundstate wave function $|\psi_g\rangle$.
 The reduced density matrices $\rho_{A/B}$
 are
 obtained from the full density matrix $\rho = |\psi_g\rangle\langle\psi_g|$
 by tracing out the degrees of freedom of the rest of the
 lattice-blocks $A$ or $B$, i.e., $\rho_{A/B} = \mathrm{Tr}_{A^c/B^c} \, \rho$.
 Thus also $\rho_{A \cup B} = \mathrm{Tr}_{(A \cup B)^c} \, \rho$.

 Based on our iMPS groundstate wavefunctions,
 we first consider the mutual information $I_q(A:B)$ between the two lattice-blocks of
 $\ell$ contiguous sites equally partitioned in a lattice-block of $2\ell$ contiguous sites,
 i.e., $\ell_A=\ell_B=\ell$ in Fig. \ref{figure1} (a).
 For the broken symmetry phases, i.e., $\lambda < \lambda_c$,
 if one chooses a random state as a reference state,
 one can detect $q$ degenerate groundstates by using the quantum fidelity \cite{su}.
 All $q$ degenerate groundstates
 give the same block-block mutual informations.
 In Figs.~\ref{fig2} (a) $q=2$, (b) $q=3$, (c) $ q=4$, and (d) $q=5$,
 we plot the mutual informations $I_q(\lambda)$
 as a function of the transverse field $\lambda$
 for various sizes of lattice-blocks $\ell$.
 One can notice that
 compared for smaller lattice-block size,
 the block-block mutual information for bigger lattice-block size has a bigger value.
 Furthermore,
 all the block-block mutual informations exhibit a singular behavior
 for various sizes of lattice-blocks $\ell$.
 The singular points correspond to the phase transition points $\lambda_c =1 $.
 Consequently, block-block mutual information can detect quantum phase transition,
 which is manifested by appearing of singular behaviors.

\section{Central charges and quantum mutual information}
  As was discussed in the previous section,
  the mutual information $I_q(\lambda)$ between two lattice-blocks
  exhibits a singular behavior at the critical points,
  which implies that block-block mutual information can capture
  quantum phase transitions for spontaneous symmetry breaking.
 When the infinite one-dimensional chain becomes a critical system,
 its universality class
 can be identified by calculating the central charge $c$ that is the main feature of
 the conformal field theory for a critical system.
 Actually,
 for the ground state $|\psi_g\rangle$ of an infinite one-dimensional critical system
 and a large block of length $\ell$, conformal field theory predicts the universal scaling of the entanglement entropy \cite{Holzhey,Vidal03-2,Latorre04,Calabrese04,Jin,Korepin,Laflorencie,Ryu,Ryu2}
 such as
\begin{equation}
   S(\ell) = \frac{c}{3} \log \ell + c',
   \label{EE}
\end{equation}
  where $c'$ is a nonuniversal constant.
 Such a logarithmic scaling of the von Neumann entropy
 can be confirmed by using our iMPS approach.
 Our numerical estimates of central charges are in excellent agreement
 with the exact values at the critical points (see Appendix A).

      \begin{figure}
      \includegraphics[width=0.5\textwidth]{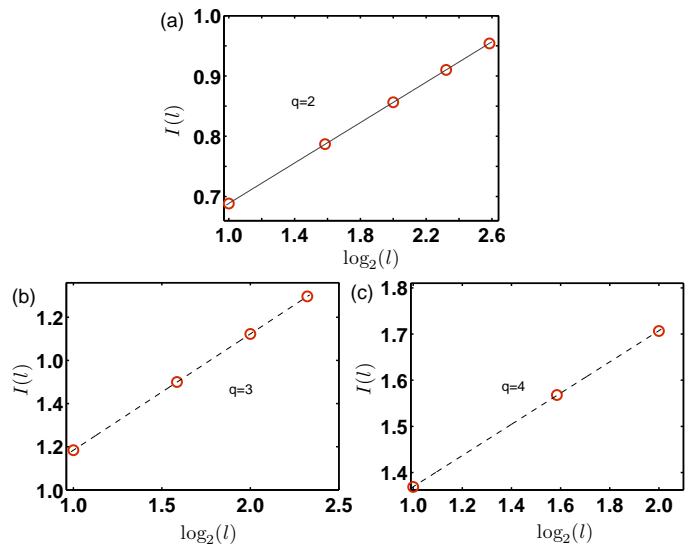}
      \caption{(color online) Mutual information $I(\ell)$
      as a function of the block length $\ell_A=\ell_B =\ell$
      at the critical point $\lambda=\lambda_{c}$ for quantum Potts chains
      with (a) $q=2$, (b) $q=3$, and (c) $ q=4$.
      The lines are the numerical fitting functions
      $I_q(\ell) = a_q\log_{2}\ell + b_q$ with the numerical coefficients $a_q$ and $b_q$.
      The detailed discussions are in the text.}
      \label{fig3}
      \end{figure}

      \begin{table}[b]
      \renewcommand\arraystretch{2}
      \caption{Numerical central charges $c$ estimated from the quantum mutual information (MI)
      and the von Neumann entanglement entropy (EE) (Appdenix A)
      at the critical point $\lambda = \lambda_c$ for $q$-state quantum Potts chains}
      \begin{tabular}{cccc}
      \hline \hline
        & $q=2 $ & $q=3$ & $q=4$   \\
      \hline
       $c$  & 1/2 & 4/5 & 1  \\
      \hline
       $c$ (MI) & 0.503(2) & 0.808(4) & 1.01(6) \\
             \hline
       $c$ (EE) & 0.5007(3) & 0.800(2) & 1.00(4) \\
      \hline\hline
      \end{tabular}
      \label{table1}
      \end{table}
%

 On varying the sizes of lattice-blocks $\ell$,
 at the critical point,
 let us then consider the mutual information $I(\ell)$ of the system in Fig. \ref{figure1} (a).
 We calculate the mutual information $I(\ell)$ as a function of the size
 of lattice-blocks $\ell$ in Fig \ref{fig3}.
 Similar to the von Neumann entanglement entropy in Eq. (\ref{EE}),
 the von Neumann mutual information $I(\ell)$ exhibits a logarithmic increment
 as the size of lattice-block  $\ell$ increases.
 Also, a similar logarithmic scaling behavior of Shannon mutual information
 with two blocks $A$ and $B$ of sizes $\ell$ and $L-\ell$
 has been conjectured for periodic chains in the ground state
 in Refs. ~\cite{Alcaraz1, Stephan14, Alcaraz2}, i.e.,
 $I_{sh}(\ell,L) = c/4 \ln[L/\pi \sin(\pi \ell/L)] + \gamma_I$
 based on the Shanon entropy in the scaling regime $(\ell, L \gg 1)$,
 where $L$ is the system size and $\ell$ is the subsystem size, and $\gamma_I$
 is the nonuniversal constant.
 In order to clarify the logarithmic behaviors of our mutual informations $I(A:B)$,
 thus we perform a numerical best fit with the fitting function
 \begin{equation}
 I_q(\ell) = a_q \log_{2}\ell + b_q,
 \end{equation}
 where $a_q$ and $b_q$ are numerical coefficients.
 The numerical fitting coefficients are given as
 (i) $a_2=0.1676(7)$ and $b_2=0.521(1)$ for $q=2$,
 (ii) $a_3=0.269(1)$ and $b_3=0.823(2)$ for $q=3$, and
 (iii) $a_4 =0.34(2)$ and $b_4=1.03(3)$ for $q=4$.
 Actually, the coefficient of the logarithm in the entanglement entropy in Eq. (\ref{EE}) is
 dependent on the central charge $c$.
 For comparison with the entanglement entropy in Eq. (\ref{EE}),
 we then consider the coefficients $3 a$, i.e.,
 (i) $3 a_2 = 0.503(2)$  for $q=2$,
 (ii) $3 a_3 = 0.808(4)$ for $q=3$, and
 (iii) $3 a_4 = 1.01(6)$ for $q=4$.
 One can notice that if one assumes that similar to the entanglement entropy
 in Eq. (\ref{EE}),
 the proportional coefficient $a_q$ corresponds to a central charge via $c_q = 3a_q$,
 our results of the central charges $c_q$ obtained from the mutual information $I(\ell)$
 are very close to the exact results
 $c=0.5$, $c=4/5$ and $c=1$ for $q=2$, $3$, $4$, respectively.
 In Table \ref{table1}, the estimates of central charges $c$ obtained from the mutual information $I(A:B)$
 are summarized for comparison with those values from the von Neumann block entropy (Appendix A).
 Consequently, it shows that
 the mutual information $I(\ell)$ may have a universal scaling behavior
 such as $I(\ell) \sim \frac{c}{3} \log \ell$ for the two lattice-blocks shown in
 Fig. \ref{figure1} (a) in critical one-dimensional infinite lattice systems.

  \section{Scaling behavior of block-block mutual information}
 For strongly correlated systems,
 a characteristic scaling behavior of conventional
 (two-point) spatial correlation can quantify their property.
 Similarly, it has
 been proposed to use the mutual information which quantifies
 the total amount of correlations including classical and quantum correlations
 shared between two parties.
 A scaling behavior of two-point (site) spatial mutual information
 has been studied and discussed its universality ~\cite{Dai1} in one-dimensional critical systems.
 In this section we study spatial behaviors of the mutual information $I(A:B)$ between the two lattice-blocks with
 $\ell_A = \ell_B = \ell$ at the critical point.
 The effects of the lattice-block size on the behaviors of mutual information will be investigated.
 We will then consider various sizes of lattice-blocks in investigating scaling behaviors of block-block mutual information.
 The detailed behaviors of  mutual information will be discussed for
 quantum $q$-state Potts model in Subsection \ref{subsection1}.
 For
 the transverse-field spin-$1/2$ XY model,
 we present a summary of scaling of mutual information  in Subsection \ref{subsection2}.
 and the detailed discussion in Appendix \ref{appB}.

 \begin{figure}
 \includegraphics[width=0.5\textwidth]{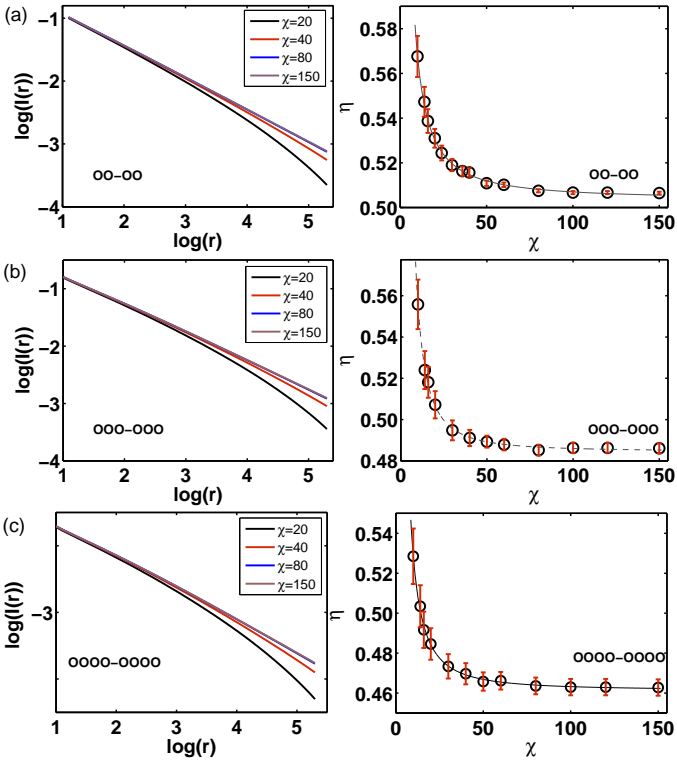}
 \caption{(color online)
  Mutual information $I(r)$ as a function of the lattice distance $r=|i-j|$
  for various truncation dimensions (left)
  and  mutual information exponent $\eta^I(\chi)$
  as a function of truncation dimension $\chi$ (right) with the block lengths
  $\ell_A=\ell_B=\ell$, i.e., (a) $\ell=2$, (b) $\ell = 3$, and (c) $\ell=4$
  for Ising chain ($q = 2$).
  Mutual information exponent $\eta^I(\chi)$ (right) is extracted from the numerical fitting of the mutual information $I(\chi)$ (left) with the fitting function  $\log I(r)=\eta^{I} \log r + a_0$
  for the power-law decaying part. The detailed discussions are in the text.}
     \label{fig4}
     \end{figure}

\subsection{Block-block mutual information critical exponent $\eta^{I}$ for quantum $q$-state Potts model}
\label{subsection1}
 Let us first consider quantum $q$-state Potts model.
 As shown in Fig. \ref{figure1} (b), when the two lattice-blocks are separated
 from each other as the distance $r$, we discuss mutual information between the two lattice-blocks.
 For $q=2$ (Ising chain), we plot the mutual information $I(r)$ as a function of the lattice distance $r$ in the left of Fig. \ref{fig4}.
 For given truncation dimensions,
 the plots show that the mutual information decreases as the lattice distance $r$ increases.
 With bigger truncation dimension, the linear region of the log-log plot becomes longer
 and the slope of the linear region seems to be readily saturated for the truncation
 dimension $\chi= 150$ in the left of Fig. \ref{fig4}.
 This tendency implies that
 similar to the power-law behavior of the von Neumann entropy in \cite{Calabrese10,Calabrese18},
 the mutual information undergoes a power-law decay to zero if the truncation dimension $\chi$ increases to the thermodynamic limit.
 Then the mutual information seems to decay linearly to zero, i.e., $I(A:B) \rightarrow 0$
 as $r \rightarrow \infty$.
 This shows that for very large separation of the two lattice-blocks,
 $S_{A \cup B} \simeq S_A + S_B$.
 Such behaviors of the mutual information $I(A:B)$ can be observed
 for all the sizes of lattice-blocks, i.e.,
 in Figs. \ref{fig4} (a) $\ell = 2$, (b) $\ell = 3$ and (c) $\ell = 4$.

 To confirm the power-law decay of mutual information $I(r)$,
 we perform a numerical fit for the linear region of mutual information $I(r)$
 with the fitting function, $\log I(r) =\eta^{I} \log r + a_0$,
 where $\eta^I$ corresponds to an exponent of power-law decay and $a_0$ is a fitting constant.
 In the right of Fig. \ref{fig4}, we plot the slopes $\eta^{I} (\chi)$ of the linear regions
 as a function of the truncation dimension $\chi$ with the fitting error bars.
 The $\eta^I$ decreases monotonically to a saturated value
 as the truncation dimension $\chi$ increases.
 In order to obtain the exponent $\eta^I_\infty$ of mutual information in the thermodynamic limit $\chi \rightarrow \infty$,
 we extrapolate the exponents for various truncation dimension in the right of Fig.~\ref{fig4}.
 The extrapolation functions are employed as a form of $\eta^I(\chi)=\eta^I_{0}\chi^\alpha+\eta^I_{\infty}$.
 The numerical estimates of the critical exponent $\eta^I_{\infty}$
 of mutual information in the thermodynamic limit are given as
 (a) $\eta^I_{0}=1.2(2)$, $\alpha=-1.26(9)$ and $\eta^I_{\infty}=0.503(1)$ for $\ell=2$,
 (b) $\eta^I_{0}=3.5(9)$, $\alpha=-1.7(1)$ and $\eta^I_{\infty}=0.485(1)$ for $\ell=3$, and
 (c) $\eta^I_{0}=2.4(8)$, $\alpha=-1.6(1)$ and $\eta^I_{\infty}=0.461(2)$ for $\ell=4$.
 For bigger size of lattice-block, the critical exponent $\eta^I_\infty$ becomes smaller.
 This means that
 the larger the lattice-blocks are, the slower the block-block information decays to zero as the lattice distance $r$ increases.
 In contrast to the mutual information,
 the von Neumann entropy was shown to be scaled as $S_{AB} \sim r^{-1/2}$ independent on
 the block sizes in \cite{Calabrese10,Calabrese18}.

 \begin{figure}
 \includegraphics[width=0.5\textwidth]{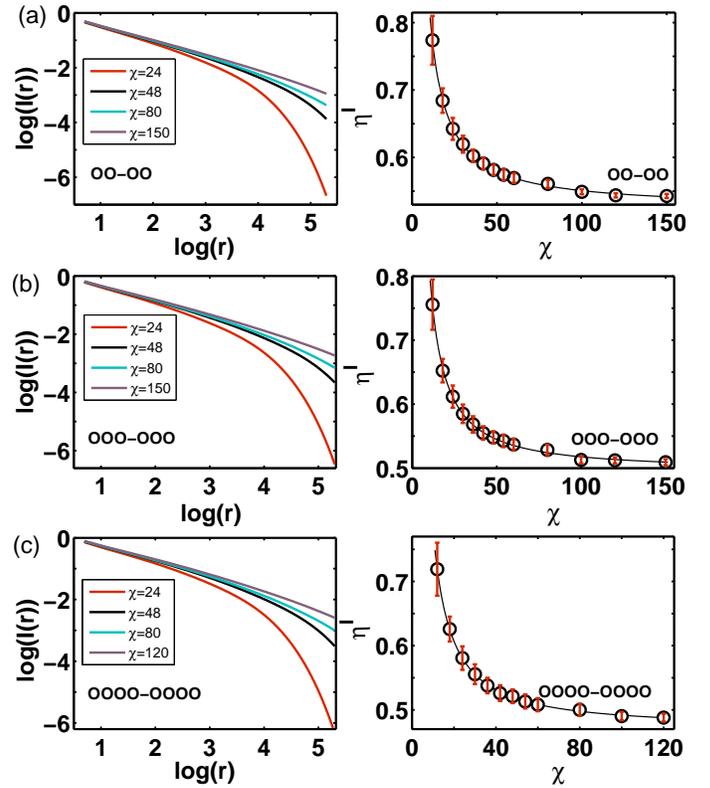}
 \caption{(color online)  Mutual information $I(r)$ as a function of the lattice distance $r=|i-j|$
  for various truncation dimensions (left)
  and  mutual information exponent $\eta^I(\chi)$
  as a function of truncation dimension $\chi$ (right) with the block lengths
  $\ell_A=\ell_B=\ell$, i.e., (a) $\ell=2$, (b) $\ell = 3$, and (c) $\ell=4$
  for three-state Potts chain ($q = 3$).
  Mutual information exponent $\eta^I(\chi)$ (right) is extracted
  from the numerical fitting of the mutual information $I(\chi)$ (left) with the fitting function  $\log I(r)=\eta^{I} \log r + a_0$
  for the power-law decaying part. The detailed discussions are in the text.}
      \label{fig5}
      \end{figure}

 For three-state Potts model ($q=3$),
 in the left of Fig.~\ref{fig5}, we display the block-block mutual information $I(r)$
 as a function
 of the lattice distance $r$ for various truncation dimension $\chi$.
 In Figs.~\ref{fig5}(a)-(c), the sizes of lattice-blocks  are chosen respectively as (a) $\ell = 2$, (b) $\ell=3$,
 and (c) $\ell=4$.
 For given truncation dimensions,
 the plots show that the mutual information decreases monotonically
 as the lattice distance $r$ between the two lattice-blocks increases.
 Similar to the case of $q=2$,
 as the truncation dimension $\chi$ increases,
 the linear region of the plot becomes longer.
 For $\chi= 150$ in the left of Fig. \ref{fig5},
 the slope of the linear region seems to be almost straight in the range of the plot.
 Regardless of the size of lattice-block $\ell$,
 such similar behaviors of mutual information $I(r)$ are noticeable.
 To analyze a characteristic behavior of mutual information,
 we adapt the approach for $q=2$.
 Using the fitting function, $\log I(r) =\eta^{I} \log r + a_0$,
 we perform the numerical fit for the linear region of mutual information $I(r)$
 in the right of Fig. \ref{fig5}.
 The slopes $\eta^{I} (\chi)$ of the linear regions
 are plotted as a function of the truncation dimension $\chi$ with the fitting error bars,
 which shows the monotonic decrement of $\eta^I$ with the increment of truncation dimension $\chi$.
 The extrapolation of the exponents is performed with the function, $\eta^I(\chi)=\eta^I_{0}\chi^\alpha+\eta^I_{\infty}$.
 We estimate the critical exponent $\eta^I_{\infty}$
 of mutual information in the thermodynamic limit as
 (a) $\eta^I_{0}=3.5(4)$, $\alpha=-1.07(5)$ and $\eta^I_{\infty}=0.525(4)$  for $\ell=2$,
 (b) $\eta^I_{0}=4.7(9)$, $\alpha=-1.16(8)$ and $\eta^I_{\infty}=0.496(6)$  for $\ell=3$, and
 (c) $\eta^I_{0}=4.7(5)$, $\alpha=-1.19(5)$ and $\eta^I_{\infty}=0.472(3)$ for $\ell=4$.
 For $q=3$,
 the estimates of the critical exponent $\eta^I_\infty(\ell)$
 show that as the size of lattice-block $\ell$ becomes bigger,
 the $\eta^I_\infty (\ell)$ becomes smaller.
 Consequently, similar to the Ising chain for $q=2$,
 the mutual information $I(r)$ for $q=3$ follows an asymptotic power-law scaling
 with a different scaling exponent.

 \begin{figure}
 \includegraphics[width=0.5\textwidth]{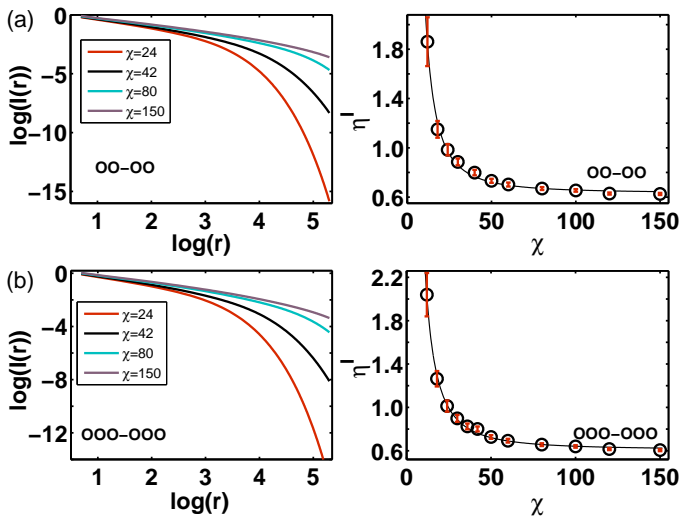}
 \caption{(color online)  Mutual information $I(r)$ as a function of the lattice distance $r=|i-j|$
  for various truncation dimensions (left)
  and  mutual information exponent $\eta^I(\chi)$
  as a function of truncation dimension $\chi$ (right) with the block lengths
  $\ell_A=\ell_B=\ell$, i.e., (a) $\ell=2$ and (b) $\ell = 3$
  for four-state Potts chain ($q = 4$).
  Mutual information exponent $\eta^I(\chi)$ (right) is extracted from the numerical fitting
  of the mutual information $I(\chi)$ (left) with the fitting function  $\log I(r)=\eta^{I} \log r + a_0$
  for the power-law decaying part. The detailed discussions are in the text. }
  \label{fig6}
  \end{figure}

 The block-block mutual information $I(r)$ reveals a very similar behavior for $q=2$ and $q=3$.
 The critical exponents seem distinguishable each other
 for a given lattice-block size.
 In order to clarify the criticality,
 the critical point of
 the four-state quantum Potts model ($q=4$)
 belonging to another universality class needs to be considered.
 Let us then consider four-state quantum Potts model ($q=4$).
 We plot the mutual information $I(r)$ for the sizes of lattice-blocks (a) $\ell = 2$ and (b) $\ell=3$
 in Fig.~\ref{fig6}.
 Similar to the cases of $q=2$ and $q=3$,
 the mutual information $I(r)$ for $q=4$ exhibit
 a power-law decaying tendency as the truncation dimension $\chi$ increases.
 By using the same numerical method for $q=2$ and $q=3$,
 we estimate the critical exponents of the mutual information $I(r)$
 for the block lengths $\ell=2$ and $\ell=3$ in the right of Fig.~\ref{fig6}.
 Performing the extrapolation with the fitting function
 $\eta^I(\chi)=\eta^I_{0}\chi^{\alpha}+\eta^I_{\infty}$,
 we get the fitting results as
 (a) $\eta^I_{0}=115(73)$, $\alpha=-1.8(3)$ and $\eta^I_{\infty}=0.63(4)$ for $\ell=2$ and
 (b) $\eta^I_{0}=118(43)$, $\alpha=-1.8(1)$ and $\eta^I_{\infty}=0.61(3)$ for $\ell=3$.
 As expected from the cases of $q=2$ and $q=3$, the exponent $\eta^I_\infty$
 is smaller for $\ell=3$ than for $\ell=2$.
 For a given size of lattice-blocks, the exponent $\eta^I_\infty$ has a distinguishable value from those of $q=2$ and $q=3$.
%
      \begin{table}[b]
    \renewcommand\arraystretch{2}
      \caption{Critical exponents $\eta^I_\infty (q,\ell) $
      of block-block mutual information $I(A:B)$ for various lattice-block sizes $\ell_A=\ell_B=\ell$ at the critical
      points for one-dimensional quantum $q$-state Potts model.}
      \begin{tabular}{c|ccc}
      \hline
      \hline
      $\eta^I_\infty (q,\ell)$ & $\ell=2$ & $\ell=3$ & $\ell=4$ \\
      \hline
       $q=2$ & 0.503(1) & 0.485(1) & 0.461(2) \\
      \hline
      $q=3$  & 0.525(4) &0.496(6) & 0.472(3) \\
      \hline
      $q=4$ & 0.63(4) & 0.61(3) &  \\
      \hline
      \hline
      \end{tabular}
      \label{table2}
      \end{table}

 Thus, for comparison, we summarize our numerical estimates
 of critical exponents $\eta^{I}_\infty$  at the critical points
 for one-dimensional quantum $q$-state Potts model in Table \ref{table2}.
 It is shown that for all of $q=2$, $3$, and $4$,
 the block-block mutual informations $I(r)$ undergo
 an asymptotic power-law scaling behavior at the critical points
 and the critical exponents become smaller for bigger lattice-blocks.
 These can be a characteristic feature of block-block mutual information
 for one-dimensional critical systems.
 Depending on $q$, i.e., universality class,
 the critical exponents $\eta^I (q,\ell)$
 seem to be given in a different value each other.

 \subsection{Block-block mutual information critical exponent $\eta^{I}$ for transverse field spin-$1/2$ XY model}
 \label{subsection2}
 In order to clarify more about universal feature of the algebraic decay
 of block-block mutual information in one-dimensional critical systems,
 we consider the transverse-field spin-$1/2$ XY model
 \cite{Lieb61,Katsura,Lieb66,Pfeuty,Damle96,Bunder99} described by the
 Hamiltonian
 \begin{equation}
 H_{XY} = -\sum_{i=-\infty}^{\infty}
      \left[ \left(\frac{1+\gamma}{2}\right) \sigma^{x}_{i}\sigma^{x}_{i+1}
       + \left(\frac{1-\gamma}{2}\right) \sigma^{y}_{i}\sigma^{y}_{i+1}
       + h\, \sigma^{z}_{i} \right],
 \label{XYHam}
 \end{equation}
 where $\sigma^{x,y,z}$ are the Pauli spin operators.
 This model has two parameters, i.e.,
 the anisotropy interaction parameter $\gamma$ and the transverse magnetic field $h$.
 As is known, the transverse-field spin-1/2 XY model has two critical lines, i.e,
 (i) the Ising transition lines with the central charge $c=1/2$ for $\gamma \neq 0$ and $h =\pm 1$,
 and (ii) the anisotropy transition line with the central charge $c = 1$ for $\gamma=0$ and $-1 < h < 1$.
 For $\gamma=1$, the XY model reduces to the Ising Hamiltonian for $q=2$ in Eq. (\ref{qHam}). Thus, in terms of the two parameters,
 $(\gamma,h)=(1.0,1.0)$ corresponds to the critical point of the Ising Hamiltonian.
 The block-block mutual information at $(\gamma,h)=(1.0,1.0)$ has been studied in Fig. (\ref{fig4}) in Subsection \ref{subsection1}.
 For comparison with $(\gamma,h)=(1.0, 1.0)$,
 we choose $(\gamma,h)=(0.5, 1.0)$ on the Ising transition line belonging to the same university class, i.e., the Ising universality class.
 Also another two parameter sets $(\gamma, h) = (0.0, 0.0)$ and $(0.0, 0.5)$
 are chosen on the anisotropy transition line belonging to the Gaussian university class.

      \begin{table}[b]
      \renewcommand\arraystretch{2}
      \caption{Critical exponents $\eta^I_\infty(\gamma,h)$
      of block-block mutual information $I(A:B)$ for various lattice-block sizes
      $\ell_A=\ell_B=\ell$ on the two critical
      lines of one-dimensional quantum transverse-field spin-$1/2$ XY model.
      For comparison, the estimates for Ising model are from Table \ref{table2} for $q=2$.}
      \begin{tabular}{c|ccc}
      \hline
      \hline
     $\eta^I_\infty(\gamma, h) $ & $\ell=2$ & $\ell=3$ & $\ell=4$ \\
      \hline
      \begin{minipage}{2.5cm} XY $(\gamma=0.0,h=0.0)$ \end{minipage}
       & 0.999(4) & 0.970(8) &0.926(5) \\
      \hline
      \begin{minipage}{2.5cm} XY $(\gamma=0.0,h=0.5)$ \end{minipage}
       &1.008(4) & 0.972(4) & 0.922(6)\\
      \hline
      \begin{minipage}{2.5cm} XY $(\gamma=0.5,h=1.0)$ \end{minipage}
       & 0.501(6) & 0.483(3) & 0.452(6) \\
      \hline
      \begin{minipage}{2.5cm} Ising ($q=2$) $(\gamma=1.0,h=1.0)$ \end{minipage}
       & 0.503(1) & 0.485(1) & 0.461(2) \\
      \hline
      \hline
      \end{tabular}
      \label{table3}
      \end{table}

%

 For the three parameter sets, we calculate the mutual information $I(A:B)$
 with the lattice-blocks $\ell_A=\ell_B=\ell=2$, $3$, and $4$.
 Similar to the one-dimensional $q$-state Potts model, the block-block mutual informations exhibit similar power-law decaying behaviors (see the details in Appendix \ref{appB}).
 In Table \ref{table3},
 their critical exponents are estimated and summarized with the case
 of the Ising critical point $(q=2)$.
 Table \ref{table3} shows clearly that
 the critical exponent $\eta^I_\infty(\gamma,h)$ decreases as the size of lattice-block $\ell$ increases.
 For each given lattice-block $\ell$, the two critical exponents $\eta^I(\gamma,h)$ of the mutual information $I(r)$ at $(\gamma, h) = (1.0, 1.0)$ and $(0.5, 1.0)$
 display a very close value each other.
 The two critical exponents at $(\gamma, h) = (0.0, 0.0)$ and $(0.0, 0.5)$ also have a very close value each other.
 Moreover, it is shown that in accordance with the Ising universality class or  the Gaussian universality class, the values of the critical exponents can be distinguishable each other
 for a given size of lattice-block.

\section{Summary}

The block-block mutual information defined by the von Neumann entropies
has been numerically investigated in the one-dimensional $q$-state quantum Potts model
and the transverse-field spin-$1/2$ XY model.
In order to calculate the reduced density matrices for the mutual information,
the groundstate wavefunction of the infinite-size lattice chain is obtained by using
the iTEBD algorithm in the iMPS representation.
We first considered the mutual information $I_q(A:B)$ between the two blocks of
$\ell$ contiguous sites equally partitioned in a block of $2\ell$ contiguous sites,
i.e., for $\ell_A = \ell_B = \ell$.
For the spontaneous symmetry breaking
in one-dimensional $q$-state quantum Potts model,
we found that all $q$ degenerate groundstates
 give the same block-block mutual informations.
Also, the block-block mutual informations $I_q(\ell)$ exhibit
a singular behavior which indicates that
a quantum phase transition occurs at the singular point and thus can be detected by using block-block mutual information.
At the critical points,
the mutual information $I_q(\ell)$
seems to have a logarithmic leading behavior, i.e., $I_q(\ell) \sim  c_q/3 \log \ell$
and the numerical coefficients of the logarithm
are shown to be a close value of the central charge $c_q$ for each $q$ in Table \ref{table1}.

  \begin{figure}
  \includegraphics[width=0.4\textwidth]{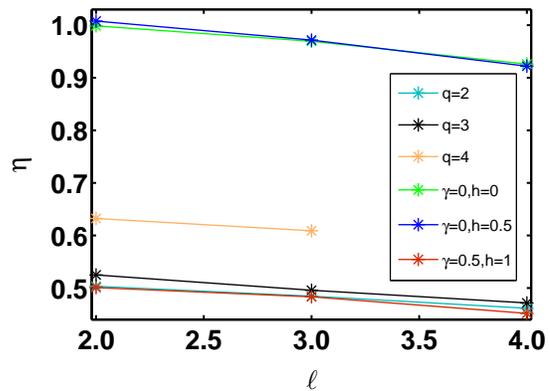}
  \caption{(color online)   Block-block mutual information exponent $\eta^I_\infty(\ell)$
  as a function of the lattice-block size $\ell$ for one-dimensional $q$-state quantum Potts model in Table \ref{table2}
  and the transverse-field spin-$1/2$ XY model in Table \ref{table3}.}
      \label{fig7}
      \end{figure}

 Next, we considered the mutual information between separated two lattice-blocks.
 In both the ordered and the disordered phases, as the distance $r$ between two lattice-blocks increases,
 the block-block mutual informations exponentially decay to zero.
 Whereas, regardless of the size of lattice-block $\ell$,
 the mutual information $I(\ell)$ is shown to undergo a power-law decay at the critical points.
 By using the extrapolation of the exponents of
 $I(\ell)$ for finite truncation dimensions,
the critical exponents $\eta^I_\infty$ in the thermodynamic limit were
estimated at the critical points.
In order to see clearly the change of critical exponent for various sizes of lattice-blocks $\ell$,
we plot the critical exponent $\eta^I_\infty$ as a function of $\ell$
in Fig. \ref{fig7} based on  Table \ref{table2}
for the $q$-state quantum Potts model and Table \ref{table3}
for the transverse-field spin-$1/2$ XY model.
It is shown clearly that the larger the size of lattice-block $\ell$ becomes,
the smaller the critical exponent $\eta^I_\infty$ becomes,
i.e., the slower the block-block mutual information decays
to zero as the distance $r$ increases.
The decreasing tendencies of the mutual informations $I(\ell)$ are similar
each other in all cases as the size of lattice-block $\ell$ increases.
For a given lattice-block size $\ell$,
the critical exponents of $I(\ell)$ for the same universality class
seem to have very close values each other in Table \ref{table3} in Subsection  \ref{subsection2}.
Whereas for different universality classes,
the values of critical exponents are different distinguishably.

\acknowledgements

    YWD is supported in part by NSFC-11805285, and
    the Fundamental Research Funds for the Central Universities (Grant No. 2019CDXYXDWL30030).
   XHC is supported by Talent Introduction Research Funds of CQWU (Grant No.
   R2019FXCo7).
    SYC acknowledges support in part from the
    National Natural Science Foundation of China (Grant Nos. 11674042,
    and 11174375).


\appendix

      \begin{figure}
      \includegraphics[width=0.5\textwidth]{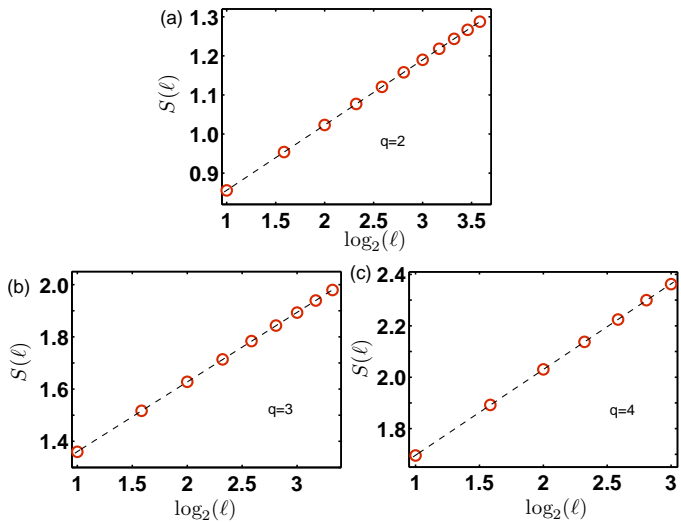}
      \caption{(color online) Von Neumann entropy $S(\ell)$
      as a function of the block length $\ell$
      at the critical point $\lambda=\lambda_{c}$ for quantum Potts chains
      with (a) $q=2$, (b) $q=3$, and $ q=4$.
      The lines are the numerical fitting functions
      $S_q(\ell) = f_q\log_{2}\ell + g_q$ with the numerical coefficients $f_q$ and $g_q$.
      The detailed discussions are in the text.}
      \label{fig8}
      \end{figure}

\section{Central charges and block entanglement entropy}
\label{appA}
 For a bipartite system, the entanglement between the two parties is defined as the
 von Neumann entropy. If one of the two parties is described by the reduced density matrix
 $\rho_\ell$ for a lattice-block of $\ell$ contiguous sites, the von Neumann entropy is given
 as $S(\rho_\ell) = - \mathrm{Tr} [ \rho_\ell \ln \rho_\ell ]$.
 When a system is in a critical regime,
 the conventional two-point correlation length diverges
 and correlations remain finite even at very large length scales.
 For one dimensional systems, conformal field theory, describes the continuum limit
 for critical systems, predicts an explicit form of the von Neumann entropy which
 relates the coefficient of logarithmic correction to the central charge of the theory.
 The entanglement entropy shows a universal behavior given in Eq. (\ref{EE}).
 In this Appendix, we calculate the von Neumann entropy for various lattice-block of $\ell$ contiguous sites
 in our iMPS groundstates.
 In Fig. \ref{fig8}, we plot the block entanglement entropy as a function of
 the lattice-block size $\ell$ for the one-dimensional (a) $q=2$, (b) $q=3$, and (c) $q=4$ state Potts models.
 As was predicted in Eq. (\ref{EE}), the entanglement entropies in Fig. \ref{fig8} exhibit
 a logarithmic scaling behavior. This fact can be manifested by performing numerical fits to extract central charges.
 With the fitting function $S_q(\ell) = f_q \log_{2}\ell + g_q$
 with the numerical coefficients $f_q$ and $g_q$.
 The numerical fitting coefficients are given as
 (a) $f_2=0.1669(1)$ and $g_2=0.6892(4)$ for $q=2$,
 (b) $f_3=0.266(8)$ and $g_3=1.094(2)$ for $q=3$, and
 (c) $f_4 =0.3334(13)$ and $g_4=1.363(3)$ for $q=4$.
 From the fitting coefficients, the central charges can be estimated as $3 f_q = c_q$, i.e.,
 (a) $3 f_2 = 0.5007(3)$  for $q=2$,
 (b) $3 f_3 = 0.800(2)$ for $q=3$, and
 (c) $3 f_4 = 1.00(4)$ for $q=4$.
 Our estimates of the central charges $c_q$ obtained from the von Neumann entropy $S_q(\ell)$
 are in excellent agreement with the exact values as was shown in Table ~\ref{table1}.
 This shows that the iMPS approach gives a reliable numerical result for the central charges.

     \begin{figure}
     \includegraphics[width=0.5\textwidth]{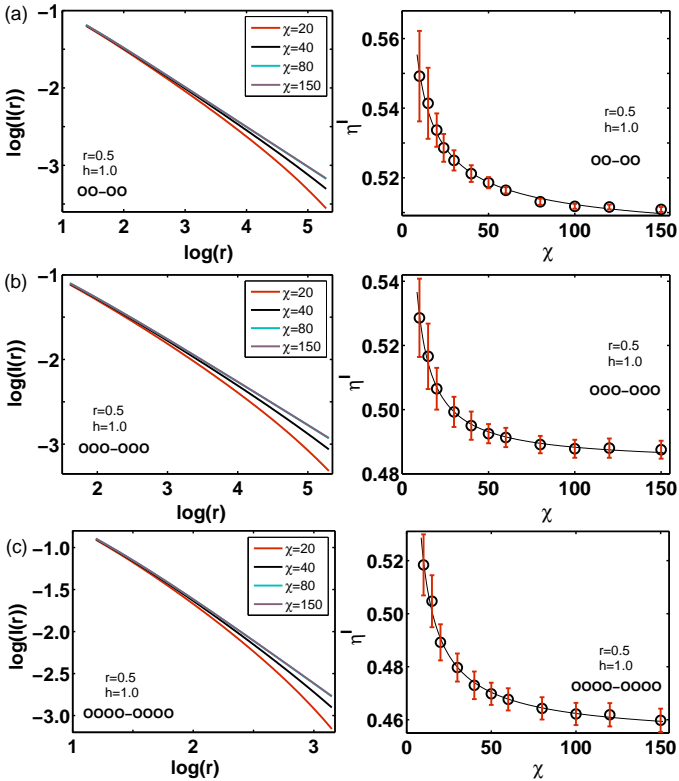}
     \caption{(color online)
 Mutual information $I(r)$ as a function of the lattice distance $r=|i-j|$
  for various truncation dimensions (left)
  and  mutual information exponent $\eta^I(\chi)$
  as a function of truncation dimension $\chi$ (right) with the block lengths
  $\ell_A=\ell_B=\ell$, i.e., (a) $\ell=2$, (b) $\ell = 3$, and (c) $\ell=4$
  for the transverse-field spin-$1/2$ XY model with $(\gamma, h) = (0.5, 1.0)$.
  Mutual information exponent $\eta^I(\chi)$ (right) is extracted
  from the numerical fitting of the mutual information $I(\chi)$ (left) with the fitting function  $\log I(r)=\eta^{I} \log r + a_0$
  for the power-law decaying part.}
     \label{fig9}
     \end{figure}

      \begin{figure}
      \includegraphics[width=0.5\textwidth]{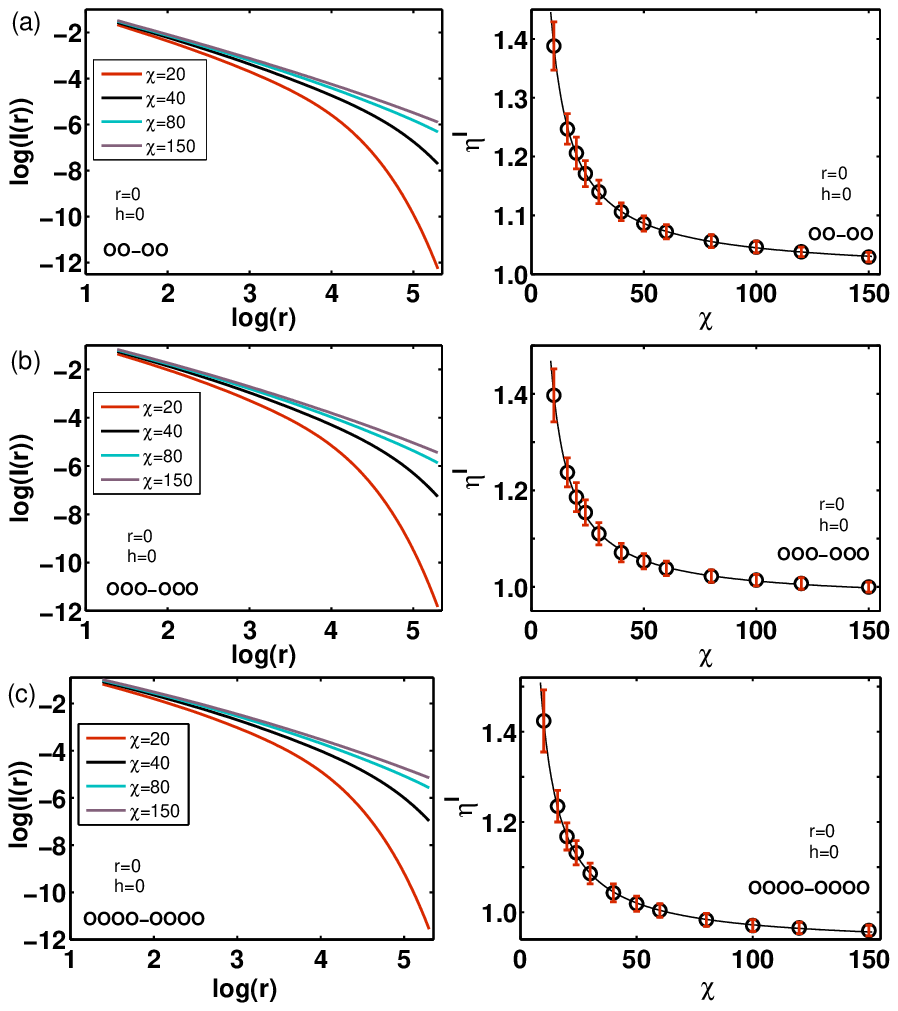}
      \caption{(color online)
      Mutual information $I(r)$ as a function of the lattice distance $r=|i-j|$
  for various truncation dimensions (left)
  and  mutual information exponent $\eta^I(\chi)$
  as a function of truncation dimension $\chi$ (right) with the block lengths
  $\ell_A=\ell_B=\ell$, i.e., (a) $\ell=2$, (b) $\ell = 3$, and (c) $\ell=4$
  for the transverse-field spin-$1/2$ XY model with $(\gamma, h) = (0.0, 0.0)$.
  Mutual information exponent $\eta^I(\chi)$ (right) is extracted
  from the numerical fitting of the mutual information $I(\chi)$ (left) with the fitting function  $\log I(r)=\eta^{I} \log r + a_0$
  for the power-law decaying part.}
      \label{fig10}
      \end{figure}
     \begin{figure}
     \includegraphics[width=0.5\textwidth]{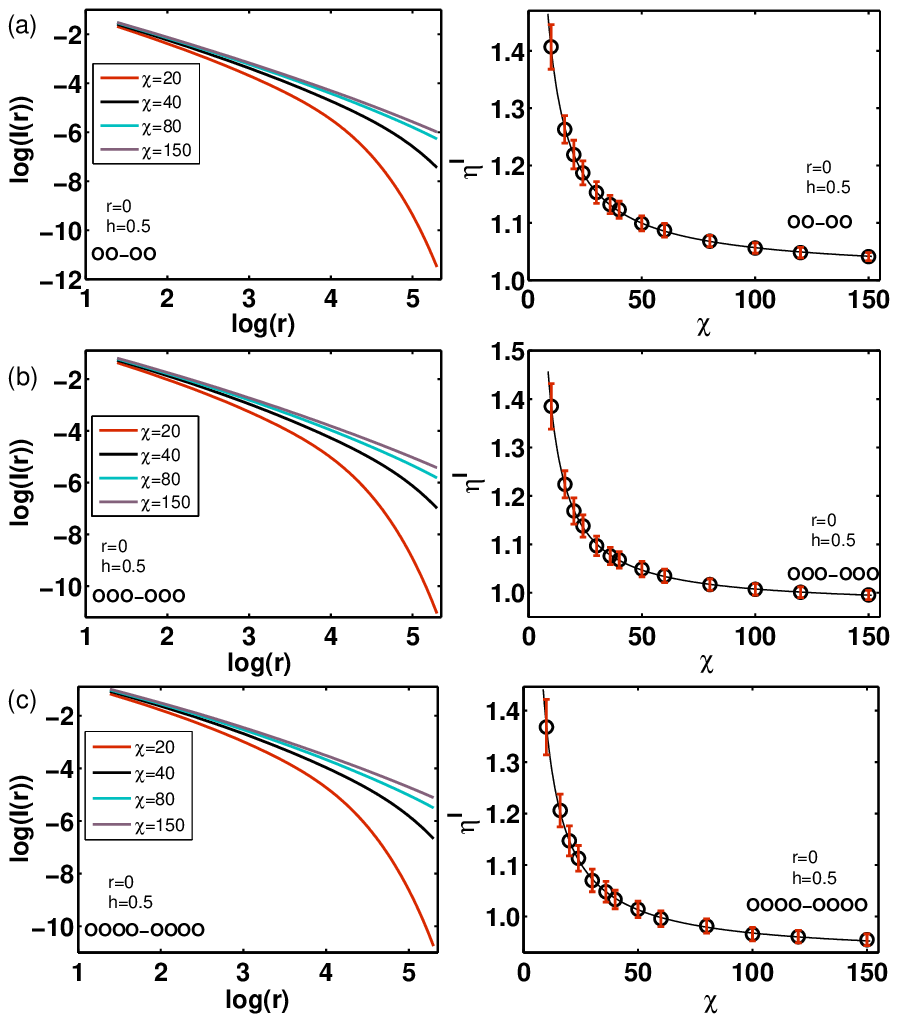}
     \caption{(color online) Mutual information $I(r)$ as a function of the lattice distance $r=|i-j|$
  for various truncation dimensions (left)
  and  mutual information exponent $\eta^I(\chi)$
  as a function of truncation dimension $\chi$ (right) with the block lengths
  $\ell_A=\ell_B=\ell$, i.e., (a) $\ell=2$, (b) $\ell = 3$, and (c) $\ell=4$
  for the transverse-field spin-$1/2$ XY model with $(\gamma, h) = (0.0, 0.5)$.
  Mutual information exponent $\eta^I(\chi)$ (right) is extracted
  from the numerical fitting of the mutual information $I(\chi)$ (left) with the fitting function  $\log I(r)=\eta^{I} \log r + a_0$
  for the power-law decaying part.}
     \label{fig11}
     \end{figure}

\section{Block-block mutual information for transverse spin-$1/2$ XY model}
\label{appB}
 In this Appendix, we will consider the model Hamiltonian in Eq. (\ref{XYHam}) to investigate the criticality of block-block mutual information $I(A:B)$. In the $\gamma$-$h$ parameter space, the four parameters are chosen for the two critical lines as
 (i) $(\gamma, h) = (1.0, 1.0)$ and $(0.5, 1.0)$ on the Ising transition line and (ii) $(\gamma, h) = (0.0, 0.0)$ and $(0.0, 0.5)$
 on the anisotropy transition line.
 For the parameters $(\gamma, h) = (1.0, 1.0)$ corresponding to the Ising model for $q=2$,
 the results of the exponents of mutual informations are displayed in Fig. \ref{fig4}.

 Figure \ref{fig9} displays the results of the mutual informations and their exponents for $(\gamma, h) = (0.5, 1.0)$.
 Overall behaviors of the mutual informations in the left of Fig. \ref{fig9}
 for $(\gamma,h) = (0.5,1.0)$ are shown to be similar with those in the left of Fig. \ref{fig4}
 for $(\gamma,h) = (1.0,1.0)$, which is that block-block mutual information $I(r)$ decays to zero algebraically
 as the distance $r$ between two blocks increases.
 In the right of Fig.~\ref{fig9}, we plot the exponents $\eta(\chi)$ of block-block mutual
 information as a function of the truncation dimension $\chi$
 for various sizes of lattice-blocks $\ell$.
 In order to get the critical exponents $\eta_{\infty}$ in the thermodynamic limit,
 the extrapolations are performed with the fitting function,
 $\eta^{I}(\chi)=\eta^I_{0}\chi^{\alpha}+\eta^I_{\infty}$ as follows:
 (a) $\eta^I_{0}=0.21(6)$, $\alpha=-0.6(2)$ and $\eta^I_{\infty}=0.501(6)$ for  $\ell=2$,
 (b) $\eta^I_{0}=0.4(1)$, $\alpha=-1.0(2)$ and $\eta^I_{\infty}=0.483(3)$ for $\ell=3$,and
 (c) $\eta^I_{0}=0.4(2)$, $\alpha=-0.8(2)$ and $\eta^I_{\infty}=0.452(6)$ for  $\ell=4$.
 One can notice that the critical exponents $\eta^I_\infty$ for $(\gamma,h)=(0.5,1.0)$ are very close values for $(\gamma,h)=(1.0,1.0)$ in
 Fig. \ref{fig4}. Note that the two parameters are in the same universality class, i.e., the Ising universality class with $c=1/2$.

  Now, let us consider other two parameters $(\gamma,h)=(0.0,0.0)$ and $(\gamma,h) = (0.0,0.5)$
  on the anisotropy transition line belonging to the Gaussian universality class with $c=1$.
  In Figs \ref{fig10} and \ref{fig11}, the results of block-block mutual informations and their exponents are
  displayed for $(\gamma,h)=(0.0,0.0)$ and $(\gamma,h) = (0.0,0.5)$, respectively.
  For the sizes of lattice-blocks (a) $\ell=2$, (b) $\ell=3$, and (c) $\ell=4$,
  as the distance between two blocks increases,
  a noticeable common behavior is an algebraic decay of block-block mutual information $I(r)$ to zero.
  In order to obtain the critical exponents of mutual information in the thermodynamic limit,
  we get the exponent $\eta^I(\chi)$ for a given truncation dimension $\chi$ from  the
  fitting function $\log(I(r))=\eta^{I} \log(r)+a_0$.
  As shown in the right of Figs. \ref{fig10} and \ref{fig11},
  the extrapolations are performed for  the critical exponents $\eta_{\infty}$
  with the function $\eta^{I}(\chi)=\eta^I_{0}\chi^{\alpha}+\eta^I_{\infty}$.
  For $(\gamma,h)=(0.0,0.0)$ in Fig. \ref{fig10},
  the numerical constants are given as
  (a) $\eta^I_{0}=3.3(2)$, $\alpha=-0.93(3)$ and $\eta^I_{\infty}=0.999(4)$ for $\ell=2$,
  (b) $\eta^I_{0}=4.3(5)$, $\alpha=-1.00(5)$ and $\eta^I_{\infty}=0.970(8)$ for $\ell=3$, and
  (c) $\eta^I_{0}=5.4(4)$, $\alpha=-1.03(3)$ and $\eta^I_{\infty}=0.926(5)$ for $\ell=4$.
  For $(\gamma,h)=(0.0,0.5)$ in Fig. \ref{fig11},
  the fitting constants are determined as   %
  (a) $\eta^I_{0}=3.3(2)$, $\alpha=-0.92(3)$ and $\eta^I_{\infty}=1.008(4)$ for $\ell=2$,
  (b) $\eta^I_{0}=4.7(3)$, $\alpha=-1.06(3)$ and $\eta^I_{\infty}=0.972(4)$ for $\ell=3$, and
  (c) $\eta^I_{0}=4.4(3)$, $\alpha=-0.99(4)$ and $\eta^I_{\infty}=0.922(6)$ for $\ell=4$.
  From these estimates, one may notice that
  for a given size of lattice-block $\ell$,
  critical exponents for $(\gamma,h)=(0.0,0.0)$ and $(\gamma,h)=(0.0,0.5)$
  are very close values each other. However, the estimate values for $(\gamma,h)=(1.0,1.0)$ and $(\gamma,h)=(0.5,1.0)$
  belonging to the Ising universality class
  are noticeably different from  those for $(\gamma,h)=(0.0,0.0)$ and $(\gamma,h)=(0.0,0.5)$ belonging to the Gaussian universality class.


\end{document}